\documentclass[a4paper,11pt]{article}
\pdfoutput=1 

\usepackage{jheppub} 
\usepackage[T1]{fontenc} 
\usepackage[dvipsnames]{xcolor}
\usepackage{amsmath}
\usepackage{amssymb}
\usepackage{amsfonts}
\usepackage{comment}
\usepackage{enumerate}
\usepackage{changes}
\usepackage{color}
\usepackage{multirow}
\usepackage{graphicx}

\definecolor{darkgray}{gray}{0.25}
\definecolor{darkgreen}{rgb}{0,0.5,0}

\newcommand{\beq}{\begin{equation}}
\newcommand{\eeq}{\end{equation}}

\def\be{\begin{equation}}
\def\ee{\end{equation}}
\def\bea{\begin{eqnarray}}
\def\eea{\end{eqnarray}}

\title{Bounding milli-magnetically charged particles with magnetars}

\author{Anson Hook,}
\author{Junwu Huang}

\affiliation{Stanford Institute for Theoretical Physics, Stanford University, Stanford, CA 94305, USA}

\emailAdd{hook@stanford.edu}
\emailAdd{curlyh@stanford.edu}

\abstract{
Milli-magnetically charged particles generically appear in scenarios with kinetic mixing.  We present model independent bounds on these particles coming from magnetars.  Schwinger pair production discharges the magnetic field of the magnetar.  Thus the existence of large magnetic fields at magnetars place strong bounds on the milli-magnetic charge to be smaller than $10^{-18}$ over a large mass range.
}

\begin{document} 
\maketitle

\flushbottom

\section{Introduction}

At low energies, particles can be described by their charge, mass and spin.  Over the years, there has been a significant effort to look for and constrain new light particles that carry a small electric charge.  Constraints on milli-electrically charged particles (mCP) have ranged from observations of astrophysical systems~\cite{Davidson:1991si, Davidson:2000hf}, cosmological constraints~\cite{Dubovsky:2003yn, Dolgov:2013una, Vogel:2013raa}, and production in colliders~\cite{Davidson:2000hf,Prinz:1998ua,CMS:2012xi}.   In this paper, we consider the slightly different case of searching for a new light particle that carries a small magnetic charge.  For other papers which have considered milli-magnetic charges, see Refs.~\cite{Brummer:2009cs, Bruemmer:2009ky, Sanchez:2011mf}.

There have also been many searches for magnetically charged particles.  A cosmological abundance of magnetically charged particles can be constrained by the neutralization of galactic magnetic fields~\cite{Parker:1970xv,Turner:1982ag,Adams:1993fj}, searches for monopoles in matter on earth~\cite{Kovalik:1986zz, Jeon:1995rf}, and direct observation~\cite{Ambrosio:2002qq, Hogan:2008sx, Detrixhe:2010xi}.  Monopoles which catalyze proton decay can be constrained with stars~\cite{Kolb:1982si, Dimopoulos:1982cz, Freese:1983hz}.  There are also searches for monopoles that involve production at colliders~\cite{Abulencia:2005hb,Abbiendi:2007ab,Kinoshita:1992wd} but are plagued by the inability to calculate cross sections due to the breakdown of perturbation theory.

In this paper, we present bounds on light milli-magnetically charged particles (mmCPs) that is independent of the microphysics that gives rise to a mmCP.  The main obstacle with providing model independent bounds on particles is that while point-like particles (e.g. the electron) can be produced with perturbative sized cross sections, extended objects (e.g. 't Hooft-Polyakov monopoles) have exponentially suppressed production cross sections~\cite{Drukier:1981fq}.  A model independent bound requires a production cross section that applies equally well to both point-like and extended objects.  An example of such a production mechanism is Schwinger pair production.  Schwinger pair production is the pair production of electric (magnetic) particles in the presence of a  large electric (magnetic) field.  The classic paper by Schwinger considered the case of pair production of point-like particles~\cite{Schwinger:1951nm} while Affleck and Manton showed that the result still holds for extended objects~\cite{Affleck:1981ag}.

While large electric fields are rare in the universe, large and relatively uniform magnetic fields occur at magnetars.  Magnetars are neutron stars with exceptionally high magnetic fields $B \sim 10^{15}$ Gauss ($10$ MeV$^2$).  Unlike the typical neutron stars, which are powered by either accretion or their spin, magnetars are instead powered by their magnetic fields (For a review of magnetars see Refs.~\cite{Harding:2006qn, Mereghetti:2008je}).  If there were light magnetically charged particles, Schwinger pair production would discharge the magnetic fields.  Requiring that the observed magnetic fields have not been discharged over the lifetime of the magnetar 
gives a very strong bound that the charge of the monopole needs to be less than $10^{-18}$.  The exact bound as a function of mass is given in Fig.~\ref{fig}.  

When discussing monopoles, there are many theoretical considerations (e.g. quantization of charge) as well as theoretical biases (e.g. monopoles must be heavy).  For the rest of the introduction, we show how kinetic mixing quite generically gives milli-magnetically charges particles. The way that quantization arguments are avoided after the $U(1)_D$ is integrated out is analogous to how quarks having a charge $e/3$ are not in conflict with the possible quantization of electric charge in units of $e$.  The prototypical example of kinetic mixing between gauge bosons of $U(1)_{\rm EM}$ and $U(1)_D$  with field strength $F$ and $F_D$ is the Lagrangian
\bea
\label{Eq: kinetic mixing}
\mathcal{L} \supset -\frac{1}{4} F^2 -\frac{1}{4} F_D^2 + \frac{\epsilon}{2} F_D F + \frac{1}{2} m_D^2 A_D^2 
\eea
where $\epsilon$ is the kinetic mixing and $m_D$ is the mass term for the dark photon $A_D$. We have given a mass to the dark photon since, as we will explain in more details later, the interactions between dark sector monopoles and our electrons vanishes when the dark photon $A_D$ is massless. The fact that kinetic mixing gives milli-magnetically charged particles is most easily seen from Maxwell's equations.  Maxwell's equations are
\bea
\label{Eq: Maxwell}
\partial_\mu F^{\mu \nu} - \epsilon \partial_\mu F_D^{\mu \nu} = e J^\mu &\qquad& \partial_\mu F_D^{\mu \nu} - \epsilon \partial_\mu F^{\mu \nu} = m_D^2 A_D^\nu + e_D J_D^\mu \nonumber \\
\partial_\mu \tilde F^{\mu \nu} = 0 &\qquad& \partial_\mu \tilde F_D^{\mu \nu} = g_D K_D^\mu
\eea
where $\tilde F^{\mu \nu} = \frac{1}{2} \epsilon_{\mu \nu \rho \sigma} F^{\rho \sigma}$, $e$ ($e_D$) are the electric gauge charges of the two $U(1)$ gauge group, while $g = 4 \pi/e$ ($g_D = 4 \pi/e_D$) are the corresponding magnetic gauge charges, $J$ ($J_D$) are the electric currents and $K$ ($K_D$) are the magnetic currents.  Maxwell's equations can be simplified by a field redefinition.  The requirement that our photon is massless fixes this field redefinition to be $A \rightarrow A + \epsilon A_D$.  To leading order in $\epsilon$, Maxwell's equations are now (see App.~\ref{App: dual} for a detailed discussion)
\bea
\label{Eq: mass basis}
\partial_\mu F^{\mu \nu} = e J^\mu &\qquad& \partial_\mu F_D^{\mu \nu} = m_D^2 A_D^\nu + e_D J_D^\mu + \epsilon e J^\mu \nonumber \\
\partial_\mu \tilde F^{\mu \nu} = - \epsilon g_D K_D^\mu &\qquad& \partial_\mu \tilde F_D^{\mu \nu} = g_D K_D^\mu
\eea
From Maxwell's equations, we see the familiar fact that our electron becomes $\epsilon$ charged under the dark sector's gauge boson.  What is perhaps less often emphasized is that the dark monopoles have obtained an $\epsilon$ magnetic charge under our photon!  If the dark sector has monopoles, then kinetic mixing naturally gives milli-magnetically charge particles~\footnote{If one finds the presence of dark monopoles disturbing, then one can use the freedom in Maxwell's equations to redefine electric and magnetic charges.  In this point of view, the dark sector has a dark electron, magnetic kinetic mixing $\epsilon F \tilde F_D$ and a magnetic mass $m_D^2 \tilde A_D^2$, where $\tilde A_D$ is the vector potential of the dual field strength satisfying $\tilde F_D = d \tilde A_D$. In this case, the dark electron will appear magnetically charged under $U(1)_{\rm EM}$ at low energies.}.  It is clear that kinetic mixing generically also generates milli-electrically charged particles.  When both mmCP and mCP are present, constraints on mmCP are typically much more stringent than constraints on mCP (see the end of Sec.~\ref{Sec: cooling} for more details).

The rest of the paper is organized as follows.  Sec.~\ref{Sec: motivate} gives a physical picture of how the milli-magnetically charged particles resulting from Eq.~\ref{Eq: mass basis} behave.  Sec.~\ref{Sec: cooling} discusses the bound from discharging the magnetic field of magnetars. Finally, we end with discussions of future directions in Sec.~\ref{Sec: discussion}. 

\section{A physical picture for milli-magnetically charged particles} \label{Sec: motivate}

In this section, we discuss how the mmCPs described by Eq.~\ref{Eq: mass basis} behave around an magnetar depending on the size of the dark photon mass $m_D$. There are two important points that govern the behavior of a mmCP.  

The first point is that mmCP are confined objects.  It has long been known that electric Higgsing is equivalent to magnetic confinement~\cite{Seiberg:1994rs, Seiberg:1994aj}.  This implies that the mmCP confine due to strings that develop between monopoles and anti-monopoles.  At close distances, just like quarks, they look like point sources.  At longer distances, the gauge fields arrange themselves to form strings between monopoles and anti-monopoles.  As there is only one scale in the problem, $m_D$, we know that the string tension is $\mathcal{O}(m_D^2)$ when $e_D$ is $O(1)$.\footnote{As we show in App.~\ref{App: pairl}, the string tension is not important as long as it is smaller than the effective magnetic field in the magnetar.}  Depending on the other forces and energy scales in the problem, one either neglects the string or needs to discuss the entire mmCP - string - anti-mmCP system (MSM). 

The second point is that the mmCP only behave like magnetically charged objects for distances $\gtrsim 1/m_D$.  
This effect can be seen in two ways.  At long distances $d \gtrsim 1/m_D$, we can integrate out the dark gauge field and from Eq.~\ref{Eq: mass basis}, it is clear that the mmCP behave as if they had an $\epsilon$ magnetic charge.  At short distances $d \lesssim 1/m_D$, we can use Eq.~\ref{Eq: Maxwell} and neglect the mass term.  Once the mass term is neglected, we can perform a different field redefinition $A_D \rightarrow A_D + \epsilon A$.  Under this field redefinition, the Maxwell's equations are instead 
\bea
\partial_\mu F^{\mu \nu} = e J^\mu + \epsilon e_D J_D^\mu + \mathcal{O}(m_D^2) &\qquad& \partial_\mu F_D^{\mu \nu} = e_D J_D^\mu + \mathcal{O}(m_D^2) \nonumber \\
\partial_\mu \tilde F^{\mu \nu} = 0 &\qquad& \partial_\mu \tilde F_D^{\mu \nu} = g_D K_D^\mu
\eea
In this basis, the mmCP and the electron interact with completely different gauge fields and thus have no interactions up to mass related propagation effects suppressed at short distances.

The other way to see that the mmCP behave like magnetically charged objects only for distances $\gtrsim 1/m_D$ is to consider the force between a current of electrons and a mmCP in the basis of Eq.~\ref{Eq: mass basis}.
A static current of electrons generates both $U(1)_{\rm EM}$ and $U(1)_D$ magnetic fields $B_\text{EM}$ and $B_D$ subject to $B_D = \epsilon B_\text{EM} e^{-m_D r}$.  Because the dark sector behaves as a superconductor, we know that, similar to the Meissner effect in a superconductor, static dark magnetic fields are screened by the mass of the $U(1)_D$ gauge boson ($m_D$). The mmCP feels a magnetic field that is
\bea
\label{Eq: Beff}
B_\text{eff} = -\epsilon B_\text{EM} + B_D = \epsilon B_\text{EM} (e^{- m_D r} - 1)
\eea
From this formula, one explicitly sees that only after the mmCP is a distance $1/m_D$ away from the electron does the mmCP appear to be millicharged.

The combination of these two facts about the mmCP explains how it evades the standard arguments for quantization of magnetic charge.  At short distances, the mmCP
behave like point-like objects without a magnetic charge.  Thus at short distances there is no violation of the quantization of magnetic charge.  At long distances, the mmCP
behave like $\epsilon$ magnetically charged objects, but also have a string attached to them due to confinement.  We thus get around the standard quantization argument by having a physical Dirac string and the fact that all finite energy objects have zero net magnetic charge.  This is completely analogous to how quarks having electric charges of $2e/3$ and $-e/3$ does not conflict with the possible quantization of electric charge in units of $e$.\footnote{Note that in either the short distance basis or the long distance basis, we find electric and magnetic charge assignments consistent with the classic Dirac charge quantization conditions (see~\cite{Brummer:2009cs, Bruemmer:2009ky} and references within).}

\section{Model Independent bounds from Magnetars} \label{Sec: cooling}

\begin{figure}
  \centering
  \includegraphics[width=0.7\textwidth]{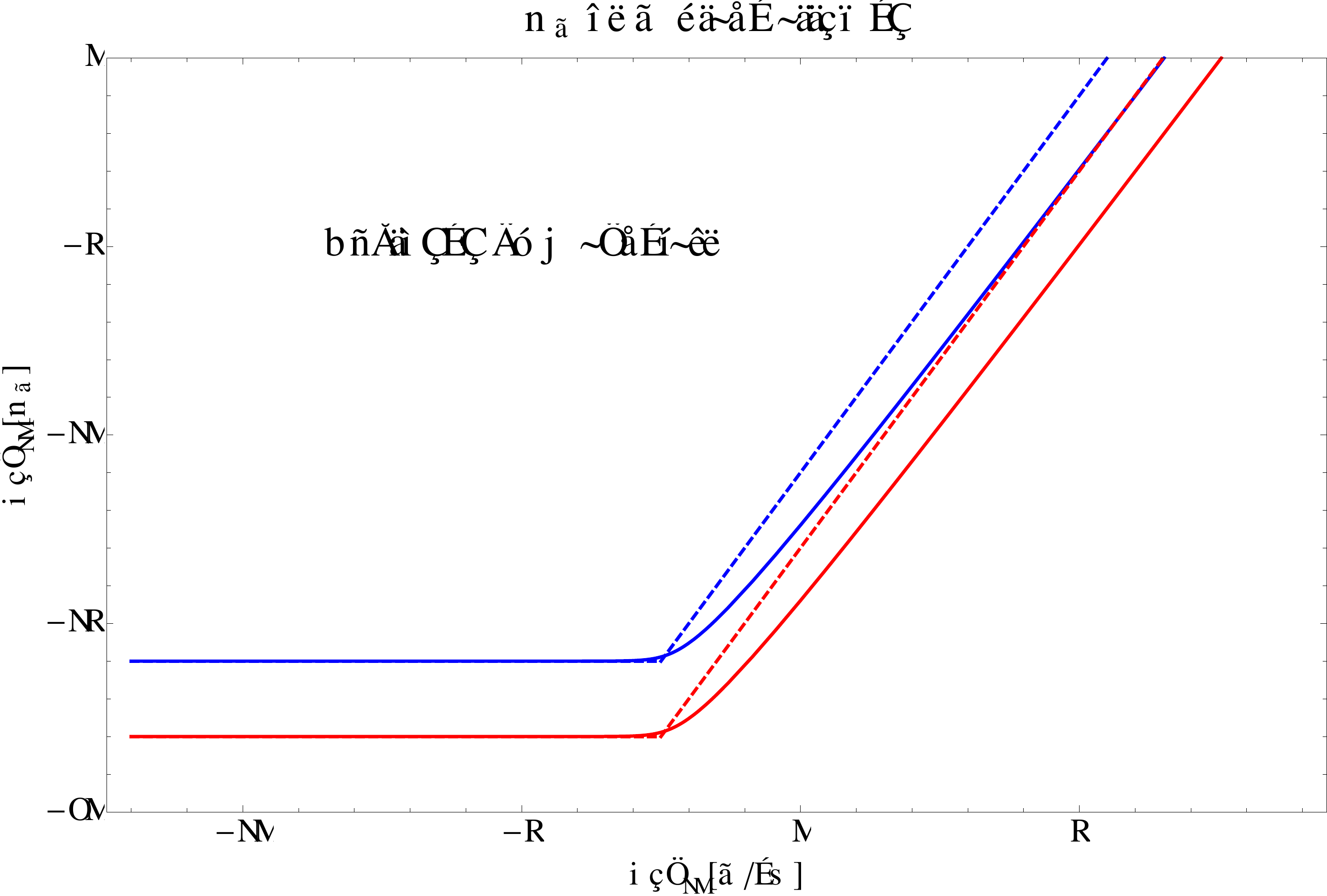}
  \caption{
    Bounds on the charge ($Q_m = \epsilon g_D/g$) of a milli-magnetically charged particle as a function of its mass m.  The red solid line (bottom) is a bound that is valid whenever the dark photon mass is in the range $m_D > 1/20$ km.  The blue solid line shows how the bound changes when $m_D = 100 /   20$ km. The dashed lines show the masses and couplings that lead to the exponent in Eq.~\ref{Eq: pair} equaling one.  The actual constraints extend a bit beyond when the exponent reaches one due to the very long lifetime of the magnetic field. 
    \label{fig}}
\end{figure}

In this section we show that Schwinger pair production of mmCPs at magnetars puts tight constraints on the magnetic charges of mmCPs.
As very little is known about magnetars, we will be extremely explicit about what we assume so that as more information arrives, it will be simple to update the bound.  Due to the uncertain nature of various quantities, we expect that the constraints we place should have order of magnitude error bars.

We assume that the magnetic field of the magnetar is a constant magnetic field of $B = 10^{15}$ Gauss and occupies a box of size $d = 20$ km, a typical field strength and size of a magnetar~\cite{Olausen:2013bpa}.  We will take the lifetime of the magnetar to be $t \gtrsim 10^4$ years, in accordance to their magnetic field powered luminosity of $L = B^2 V/2 t \approx 5 \times 10^{35}$ ergs/s.
We will be bounding a fermionic mmCP with a mass m and a milli-magnetic charge of $Q_m = \epsilon g_D/g$ measured in units of $g = 4 \pi/e$. 

The pair creation rate of mmCPs is given by~\cite{Affleck:1981ag}
\bea
\label{Eq: pair}
\frac{d N}{dV \, dt} = \frac{Q_m^2 g^2 B_\text{eff}^2}{4 \pi^3} e^{\frac{- \pi m^2}{Q_m g B_\text{eff}}} \qquad g = \frac{4 \pi}{e}
\eea
As mentioned  in Sec.~\ref{Sec: motivate}, the effective B field felt by the mmCP depends on the distance of the mmCP from what is creating the magnetic field.  We approximate the averaged magnetic field felt by the mmCP by
\bea
B_\text{eff} = B (1- e^{-m_D d})
\eea

After being produced, the mmCP are accelerated by the magnetic field such that the energy extracted from the magnetar per mmCP created is
\bea
\label{Eq: Eloss}
E_\text{loss} = g Q_m  B_\text{eff} d.
\eea
A mmCP will get accelerated to a very high boost by the large uniform magnetic field inside and around a magnetar. This result holds so long as the force on the monopole from the string is smaller than the force pulling them apart
\bea
\label{Eq: force}
m_D^2 \lesssim g Q_m  B_\text{eff}.
\eea
Eq.~\ref{Eq: Eloss} holds for the toy example of magnetic field lines parallel to the sides of the box.  A more realistic treatment will change the results by $\mathcal{O}(1)$ numbers.  Given the uncertainties in the various parameters of the magnetar, this toy model is sufficient.  Conservation of energy dictates that the energy picked up by the mmCP must come from the magnetic field. 

The energy loss from the magnetic field energy density $\frac{1}{2} B^2$ due to monopole pair-production and acceleration is
\bea
\label{Eq: energy loss}
\frac{d E}{d t \, d V} =   \frac{Q_m^2 g^2 B_\text{eff}^2}{4 \pi^3} e^{\frac{- \pi m^2}{Q_m g B_\text{eff}}}  E_\text{loss}.
\eea
Requiring that the magnetic field has survived over the course of the lifetime of the magnetar $t_{\rm life}$ is the constraint that Eq.~\ref{Eq: energy loss} is smaller than $B^2/2 t_{\rm life}$. It should be noted that the energy loss from the magnetic field does not depend on the details of how the energy leaves the system as long as they do not replenish the magnetic field energy density. The energy density can escape the magnetar in the form of boosted mmCPs for very small string tension ($m_A \rightarrow 0$), and in the form of a MSM system, which radiates its energy in the form of bremsstrahlung photons, for very large string tension ($m_A \gtrsim m$). In the intermediate region where $0 \ll m_A \ll m$, the exact form of the escaping energy density depends on detailed dynamics of the MSM system in the magnetar since the MSM system might be trapped around the magnetar for a considerate amount of time due to the interactions between the string and the charged particles inside and around the magnetar.  This energy either leaves the system in the form of mmCPs, MSMs, high energy photons, or gets reabsorbed by the magnetar in the form of thermal and non-thermal radiation, and thus will not replenish the magnetic field. Therefore, though it is very important to understand how much of the energy escaped the system and in what form in order to design dedicated ground based experiments to search for either mmCPs or strings originating from a magnetar, the constraints from the destruction of the magnetic field due to Schwinger pair production of mmCPs  are not affected.  We leave these discussions of these considerations to future work.

Milli-magnetically charged particles are described by three different parameters, $Q_m$, $m$ and $1/m_D$ describing respectively its charge, mass, and length scales beyond which it behaves as a magnetically charged object.  As long as $m_D \gtrsim 1/20$ km, then the value of $m_D$ is unimportant for the constraints resulting from Eq.~\ref{Eq: energy loss}.  This bound is shown as a solid red line in Fig.~\ref{fig}.  For $m_D \lesssim 1/20$ km, the value of $B_\text{eff}$ is sensitive to the dark photon mass.  We show how the bounds behave in this mass region for two fixed values of $m_D=1/2$ km and $m_D=5$ km.

In Fig.~\ref{fig}, there is a region where the Schwinger pair production rate in Eq.~\ref{Eq: pair} does not strictly apply.  Note that the bound flattens out in Fig.~\ref{fig}. This is because Schwinger pair production is no longer exponentially suppressed and we need to sum over higher order terms in the expansion. However, we know that the probability of pair production must be parametrically $Q_m^2 g^2 B_{\rm eff}^2$ up to $O(1)$ numbers resulting from the summation. This result is analogous to sphalerons at finite temperature where above the critical temperature the probability of sphalerons occurring is just $T^4$ as set by dimensional analysis.  Motivated by this, we use Eq.~\ref{Eq: pair} even in the regime where it is not exponentially suppressed as it is expected to be $\mathcal{O}(1)$ off from the proper expression~\footnote{In the case of an electric field in a periodic box, even the $\mathcal{O}(1)$ numbers in front of the exponential stay the same as $m \rightarrow 0$~\cite{Cohen:2008wz}.}. We find as a result
\begin{equation}
Q_m > 10^{-18} \left(\frac{20 \, {\rm km}}{d}\right)^{1/3}\left(\frac{10^4 \, {\rm years}}{t}\right)^{1/3} \left(\frac{10^{15} \,{\rm Gauss}}{B}\right)^{1/3}
\end{equation}
for light mmCP mass and $m_D > 1/20$ km.

The other concern is if the results apply when $1/m$ of the mmCP is larger than the size of the magnetar as the magnetic field is inhomogeneous on those scales.  Because the mmCP are produced with characteristic kinetic energy $\sqrt{Q_m g B_\text{eff}}$, their Compton wavelength is in fact smaller than the magnetar and there is no problem even when the mass of the mmCP vanishes~\cite{Cohen:2008wz}. Computations of Schwinger pair production in inhomogeneous fields~\cite{Kim:2000un} show that corrections to the Schwinger formula are mass suppressed.

\subsection*{Constraints on electric versus magnetic particles}

Many of the theories which give magnetically charged particles also give electrically charged particles.  For example, one might get a dark monopole by breaking an $SU(2)_D$ down to $U(1)_D$ with an adjoint Higgs field.  There are magnetically charged 't Hooft-Polyakov monopoles as well as electrically charge massive gauge bosons.
A possible concern is whether the constraints on the kinetic mixing parameter $\epsilon$ from these additional particles are stronger than those due to the mmCP.  
This question can be most simply answered by looking at Eq.~\ref{Eq: Maxwell} and Eq.~\ref{Eq: mass basis}.  It is clear from these equations that we are dealing with a mmCP with a magnetic charge $Q_m = \epsilon g_D/g$, a mCP with an electric charge $Q_e = \epsilon e_D/e$ and a dark photon with kinetic mixing $\epsilon$.
Light mmCP have $Q_m < 10^{-18}$ which translates to $\epsilon \lesssim 10^{-16}$ using only the perturbativity of $e_D$.
Right away, one can see that if the mmCP is light ($m \lesssim {\rm eV}$), then the constraint on $\epsilon$ from the mmCP is stronger than any other constraint on $\epsilon$, the strongest of which comes from red giant cooling at $\epsilon \lesssim 10^{-14}$ (for a review on the various constraints on mCPs and dark photons, see~\cite{Jaeckel:2010ni} and references within). In general, the more perturbative the dark sector is (the smaller $e_D$) the weaker the mCP bounds are and the stronger the mmCP bounds are on $\epsilon$.

\section{Discussion} \label{Sec: discussion}

In this paper, we described how milli-magnetically charged particles are constrained by the existence of the large magnetic fields at magnetars.  mmCP can be produced by Schwinger pair production and their subsequent ejection from the magnetar is a new energy loss mechanism that discharges its magnetic field.  At its strongest, this constrains the mmCP to have a magnetic charge smaller than $10^{-18}$.

As we learn more about magnetars, there may be additional interesting features to consider.  One example is if the dark photon mass arises from the Higgs mechanism rather than the Stuckleberg mechanism, there is a worry of symmetry restoration in a large magnetic field.  In the presence of a magnetic field, the mass of the dark Higgs boson is modified by
$m_H^2(B) = -m_{H}^2 + \epsilon g B$.
If the magnetic field is large enough, then symmetry may be restored in the dark sector resulting in the mmCP feeling an effectively zero magnetic field.

There are many interesting things to do with this result.  One amusing result is that magnetars provide a new non-cosmological production mechanism for magnetically charged objects, and in much of the parameter space also the dark photon.  If the milli-magnetically charged particle are being produced but do not carry off enough energy to destroy the magnetic fields, one might hope to observe them at Earth.  Detection of mmCP will be different from typical monopole searches due to their small charge. Naively, it seems unlikely that these mmCPs will be observable with ground-based experiment since the nearest observed magnetars are $\sim 10\, {\rm kpc}$ away and the mmCPs have relatively small charge to mass ratio and interactions with standard model fermions are suppressed by the dark photon mass. However, two possibilities remain. Firstly, the mmCPs from the magnetars are very boosted due to acceleration by the strong magnetic field of a magnetar ($\gamma \sim Q_m  B_{\rm eff} d/m$). Depending on the distribution of the magnetic field of the magnetar, the mmCPs from the magnetars can be confined within a jet with very small openning angles, resulting in a much enhanced flux of mmCPs once the jet sweeps through the earth. A large flux of MSM from magnetars is not inconsistent with the Parker bound~\cite{Parker:1970xv} on magnetic monopoles since the mmCPs are very weakly coupled and, in most of the parameter space, the MSM system is much smaller than the galactic scale and behaves as a magnetic dipole. Secondly, the strings that connect the mmCP pairs can have macroscopic length and are therefore much more likely to be discovered. These strings carry electromagnetic energy and can be looked for with experiments recently proposed to search for dark photon dark matter. The newly proposed experiments~\cite{Bunting:2017net} can be potentially senstive to these strings if they reach design sensitivity even when the dark photon is not the dark matter, though it depends on the masses and couplings of the monopole, the dark photon mass, details of MSM system and string loop dynamics in a strong magnetic field. We will leave these analysis to a future work.

\subsubsection*{Acknowledgments}
We thank Jeremy Mardon for collaboration at early stages of the project and many important
discussions. We thank Asimina Arvanitaki, Savas Dimopoulos, Sergei Dubovsky and Ken Van Tilburg for helpful discussions. We thank Masha Baryakhtar and Gustavo Marques-Tavares for valuable discussions and comments on the draft. A.H. and J.H. are supported by NSF Grant PHYS-1316699. A.H. is also supported by the DOE Grant DE-SC0012012.

\appendix

\section{Transformation properties of the dual photon} \label{App: dual}

In this Appendix, we describe how to obtain the transformation properties of the dual photon (which couples to monopoles) from the transformation of the photon (which couples to electrons).   As mentioned before, we are interested in the field redefinition in the presense of a $m_D^2 A_D^2$ term in the Lagrangian.
\bea
\label{Eq: field redefinition}
A \rightarrow A + \epsilon A_D \qquad A_D \rightarrow A_D
\eea
This field redefinition results in new currents
\bea
J \qquad J_D + \epsilon J
\eea
where $J$ ($J_D$ )are the currents associated with $A$ ($A_D$).  In this basis, everything charged under EM becomes milli charged under the dark gauge boson.

To see how the dual photon and the magnetic currents are changed under this field redefinition, we need to go from the electric frame to the magnetic frame.  This is done by a Legendre transformation.  The case of a single $U(1)$ gauge boson is given schematically by
\bea
\mathcal{L} = -\frac{1}{4}F^2 + \frac{1}{2}\tilde F G + K B
\eea
where $G = dB$.
Notice that we can integrate out G and arrive at the original theory but with the additional constraint that
\bea
\partial \tilde F = K
\eea
This means that K is the source of monopoles and B is the dual photon.  Integrating out F instead of G, we get a theory
\bea
\mathcal{L} = -\frac{1}{4}G^2 + K B.
\eea
We have a theory of the dual photon instead of a theory of the photon.

We now apply this trick to the previous case of two photons with kinetic mixing.
\bea
\mathcal{L} = -\frac{1}{4}F^2 -\frac{1}{4} F_D^2 + \epsilon F F_D + \frac{1}{2}\tilde F G +\frac{1}{2}\tilde F_D G_D + K B + K_D B_D
\eea
where we have suppressed $\mathcal{O}(1)$ numbers and gauge coupling constants.  $G$ and $G_D$ are the dual photons of EM and the dark $U(1)$.  We now do the field redefinition in Eq.~\ref{Eq: field redefinition}, while requiring that $G$ and $G_D$ are still Legendre polynomials.  This means that when we field redefine $F$ and $F_D$, we also do a field redefinition on $G$ and $G_D$ so that there are no mixed $\tilde F G_D$ terms.  The field redefinition of the dual photons is
\bea
B \rightarrow B \qquad B_D \rightarrow B_D - \epsilon B
\eea
The final Lagrangian is thus
\bea
\mathcal{L} = -\frac{1}{4}F^2 -\frac{1}{4}F_D^2 +  \frac{1}{2}\tilde F G + \frac{1}{2}\tilde F_D G_D + ( K - \epsilon K_D ) B + K_D B_D
\eea
Note that this means that the new magnetic currents are
\bea
K - \epsilon K_D  \qquad  K_D 
\eea
Monopoles of the dark sector become milli-magnetically charged under EM.

\section{A brief review of Schwinger Pair production}
\label{App: pairl}

In this Appendix, we extremely briefly discuss Schwinger pair production when the particles being created have a string attached between the two.
A mmCP accelerated from rest in a magnetic field has the world line
\bea
x^0(\tau) = r \sinh ( \frac{g B - F_\text{string}}{m} \tau) \qquad x^1(\tau) = r \cosh ( \frac{g B - F_\text{string}}{m} \tau)
\eea
where $F_\text{string}$ is the force on the mmCP from the string.  Bounce actions are done in euclidian space where these hyperbolic equations of motion become a circle.

The bounce action is
\bea
S &=& - \int d^4x \frac{1}{4} F_{\mu \nu}^2 + m \int d\tau \sqrt{\dot x^\mu \dot x_\mu} + m_D^2 \int d^2 \sigma \sqrt{-\det \partial_a \sigma^\mu \partial_b \sigma_\mu} \nonumber \\
&=& - \pi r^2 g B + 2 \pi r m + \pi r^2 m_D^2
\eea
The bound action is minimized at 
\bea
r = \frac{g B - m_D^2}{m} \qquad S = \frac{\pi m^2}{g B - m_D^2}
\eea
This value of the radius can also be determined separately by energy conservation.
We see that the Schwinger result is not significantly changed as long as $m_D^2 < g B$.  This inequality is satisfied as we already imposed it in Eq.~\ref{Eq: force}.

\bibliography{reference}
\bibliographystyle{JHEP}

\end{document}